\RequirePackage{xspace}

\documentclass[preprint,12pt]{elsarticle}
\usepackage{amssymb}

\usepackage{graphicx}
\usepackage{bm}
\usepackage{amsmath}
\usepackage{multirow}
\usepackage{amsbsy}

\usepackage{mathptmx}

\begin{document}

\begin{frontmatter}

\title{\boldmath  Multivariate Goodness of Fit  Procedures for Unbinned Data: An Annotated Bibliography} 

\author{G.~Palombo}
\address{ California Institute of Technology}
\ead{palombo@cacr.caltech.edu}

\begin{abstract}

Unbinned maximum likelihood is a common procedure for parameter estimation. After parameters have been estimated, it is crucial to know whether the fit model adequately describes the experimental data. Univariate Goodness of Fit procedures have been thoroughly analyzed. In multi-dimensions, Goodness of Fit test powers have rarely been studied on realistic problems. There is no definitive answer to regarding which method is better. Test performance is strictly related to specific analysis characteristics. In this work, a review of multi-variate Goodness of Fit techniques is presented.      
\end{abstract}





\end{frontmatter}


\section{Introduction}
\label{intro}

An important task in particle physics analysis is to estimate how well the fit function approximates the measured observations. Goodness-of-fit (gof) tests perform this task. The  null hypothesis $H_0$  is that the observed data follow some specified probability density. 

A large number of univariate gof methods for unbinned data exist \cite{Dagostino}. Some of them can be extended to two or three dimensions. However, they typically use an ordering scheme which is not easily extensible to many dimensions.
On the other hand,common gof tests for binned data, such as $\chi^2$ tests, can be theoretically  used in many dimensions. Unfortunately, as the number of dimensions increases, they  become quite inefficient, suffering from the ``curse of dimensionality''  \cite{Bellman}. In fact, unless the data sample is extremely large, the majority of bins are empty.
The goal of this work is to give a comprehensive review of multivariate gof techniques for unbinned data. 
The power of gof methods varies greatly depending on the characteristics of the data, alternative hypotheses, goals of the analysis, etc... There does not exist a multivariate gof test that is always better. Therefore, the problem to address is not which multivariate unbinned gof technique is best, but rather which one is the best gof technique given different analysis configurations.

In the gof procedure, a set of N sample data $ \{{\bf x}_i\}_{i=1}^N$ is observed. The null hypothesis is that the random variable $ {\bf X}$ follows a given distribution $  p {\bf(x)}$.  Given $H_0$ is true, the probability of accepting the null hypothesis is defined as the confidence level of the test ($1-\alpha_I$, where $\alpha_I$ is usually fixed to a few percent). On the other hand, given $H_0$ is false, the probability of correctly rejecting the null hypothesis ($1-\alpha_{II}$) is defined as the power of the test. In statistics, $\alpha_I$ and $\alpha_{II}$ represent Type I and Type II errors. One fixes the Type I error, and then looks for the test with the highest power.
An alternative hypothesis $H_1$ is usually a composite hypothesis. That is, it gives very little information about the distribution of the data and is typically in the form of $  {\bf X} \ensuremath{\sim}  g{\bf(x)} $, with $ g{\bf(x)} \neq p{\bf(x)}$. In the case where more is known about the alternative hypothesis, that information should be included in the test in order to increase its power.

Two-sample testing problem is strictly connected  to gof and can as well be used to perform a fit test. To test the gof, a set of events can be randomly drawn from the null hypothesis distribution and a two-sample test can be performed. The size of the hull hypothesis sample should be large, but not too large with respect to the observed events \cite{FriedmanNew}.

A useful step before performing any multivariate gof test is to use the theorem due to Rosenblatt \cite{Rosenblatt} that allows one to transform the original multidimension random vector with a given joint pdf, to a vector uniformly distributed on a multidimension unit cube. However, in multidimensions the transformation to uniformity is not unique. Different transformations lead to different conclusions. For a detailed discussion and suggestions on multivariate transformation to uniformity, see Ref. \cite{Narsky}.

In general, all multivariate gof methods described in this work do not specifically need a  transformation to uniformity. However, a transformation to the unit hypercube is necessary to make the gof test equally sensitive to all regions of the observed space. 

\section{Multivariate Tests for Multinormal Distributions}
\label{sec:1}

Multivariate normality can be tested by considering separately each marginal variable \cite{Dagostino}. Each marginal distribution is tested by itself using univariate gof procedures. The presence of at least one non-normal distribution indicates non-normality of the multivariate distribution. Information about the univariate distributions can often give a good understanding about the multivariate distribution as well. For the significance of each univariate test, D'Agostino and Stephens (1986) suggest to use $ \alpha / p$ where $p$ is the number of dimensions and $\alpha$ the desired gof test significance. For a  comprehensive review on how to combine significance levels, see Ref. \cite{Cousins}.
An obvious drawback of this approach is that normality of all marginals does not imply joint normality. This is particularly disadvantageous considering that in gof tests, unlike the majority of hypothesis tests, one usually wants to accept the null hypothesis. 

Mardia (1970) is considered to be the first to have proposed a multivariate test for normality. 
Mardia suggests to test multinormality by estimating multivariate skewness and kurtosis.  No significant skewness or kurtosis leads one to accept the null hypothesis \cite{Mardia}.

Malkovich and Afifi (1973) also suggest to use skewness and kurtosis to check for multinormality. They propose to generalize the univariate measures of skewness and kurtosis by using Roy's union-intersection principle \cite {Malkovich}. The asymptotic distribution of the statistics used by these tests can be found in Ref. \cite {Machado}. They also present a generalization of Shapiro and Wilk statistics for multivariate tests. 

Cox and Small (1978) propose to look for the linear combination of the pair of variables that maximizes the curvature of one variable when regressed on the other. The maximum curvature is the statistical test \cite{Cox}.

Andrews et al. (1972) introduce the nearest distance test  for joint normality. Firstly, points are transformed to the unit hypercube. Then, for each point, the smallest distance to another point is estimated. Finally, independence is tested between a specific transformation of the distances and the original points from which the transformation has been derived. Multiple  regression techniques are used to test for independence \cite{Andrews}.

A number of techniques have been proposed using  scaled residuals, defined by Andrews et al. (1971) as

\begin{equation}
Z_i =  S^{- \frac {1} {2}}( X_i - \overline{X}) 
\end{equation}

where $ S^{- \frac {1} {2}}$ is the symmetric square root of the covariance matrix $S$; $X_i$, for $i=1, . . . ,n$, are the $n$ observations and $\overline{X}$ is the sample mean vector.
The $Z_i$ are distributed symmetrically provided that original distribution is normal.
Andrews et. al (1971) suggest to create a vector defined as a normalized weighted sum of  $Z_i$.  Projections of the original observations onto the space defined by the vector are univariate. Therefore, a univariate test can be employed  \cite{Andrews2}.

From the scaled residuals, squared radii or angles can be derived. The squared radius is defined as

\begin{equation}
r_i^2 = {Z_i}'Z_i
\end{equation}

and $\theta_i$ is the angle made by $Z_i$ with a given line. Under the null hypothesis,the probability plots for radius and angle should be linear.
Techniques based on radius and angle are typically used to estimate bivariate distributions.

Other tests for multinormality have been proposed by Dahiya and Gurland (1973) \cite{ Dahiya}, Hensler et al. (1977) \cite{Hensler}, Csorgo (1986) \cite{Csorgo}, Mudholkar et al. (1992) \cite{Mudholkar} and Ghosh and Ruymgaart (1992) \cite{Ghosh}.

Unfortunately, a comparison among the above-described tests has never been carried out. That makes it hard to give recomendations. 
Mardia's test is often considered the most reliable test for multinormality. D'Agostino and Stephens suggest combining Mardia's test with univariate tests for marginal normality to gain  useful information about the multivariate distribution \cite{Dagostino}.     

\section{Kolmogorov-Smirnov Test}
\label{sec:2}

The Kolmogorov-Smirnov univariate test is an extremely popular gof procedure   \cite {Dagostino}. Given a random sample $ x $ of size n,  define $F(x)$ as the Cumulative Distribution Function (CDF) under $H_o$, and define the Empirical Distribution Function (EDF) as

\begin{equation}
F_n(x)=\frac {\#  \; obs. \leq x} {n},
\end{equation}

The Kolmogorov-Smirnov test estimates the maximum absolute difference between $F_n(x)$ and $F(x)$. That is,

\begin{equation}
D= \sup_x {|F_n(x) - F(x)|}.    
\end{equation}

Small values for $D$ lead one to accept the null hypothesis.
Other possible statistics can be created by just considering the maximum positive ($\sup_x {F_n(x) - F(x)}$) or negative ($\sup_x {F(x) - F_n(x)}$) difference, as well as the sum of the maximum positive and negative difference (Kuiper's test) \cite {Dagostino}.

Justel et al. (1997) propose a generalization of the  Kolmogorov-Smirnov test in high dimensions that has become the most popular one. The extension of $D$ in $p$ dimensions becomes:

\begin{equation}
D_n= \sup_{{\bf x} } {|F_n({\bf x}) - F({\bf x})|};    
\end{equation}

The estimation of $D_n$ is complicated for $p>2$. Therefore, Justel et al. suggest to approximate $D_n$ with  

\begin{equation}
\widetilde{D_n}= \sup_{{\bf x} \epsilon A} {|F_n({\bf x}) - F({\bf x})|};    
\end{equation}

where $A$ is a domain for  $F({\bf x})$. For  large n, the power of the exact and approximate tests is similar. Percentiles of these distributions can be found in Ref.  \cite {Justel}.  
The Kolmogorov-Smirnov univariate test is usually quite powerful. However, as pointed out in Ref. \cite{Dagostino, Narsky}, it is inefficient in detecting small clusterings in the data. 
 Kolmonorov-Smirnov multivariate test is not as reliable as the univariate one. 

The Cramer-von Mises EDF statistic  \cite {Dagostino} estimates the integrated quadratic discrepancy between $F_n(x)$ and $F(x)$,

\begin{equation}
Q= n \int_{- \infty}^\infty \{F_n(x)-F(x)\}^2 \psi(x) \, dF(x);    
\end{equation}  

where $\psi(x)$ is a suitable weighting function. See Ref. \cite{Dagostino} for a detailed description of different statistics derived by different choices for $\psi(x)$.
Univariate gof of fit tests based on the Cramer-von Mises statistic are widely used. However, a general and reliable extension to higher dimensions  has not yet been proposed \cite{Justel}, although the multivariate distribution of Cramer-von Mises statistic has been thoroughly studied \cite{Johnson}.
 
A major issue with the extension into the higher dimensions of EDF tests is that it leads loss of sensitivity in some parts of the observable space. In the multivariate space, some studies have shown that EDF gof test power quickly degrades \cite{Narsky,Friedman}.  

\subsection{MLV }
\label{sec:3}

Maximum Likelihood Value (MLV) uses the value of the likelihood function at its maximum to check how well the functional form approximates the observed events. A detailed description of this method can be found in the BaBar Statistics Report (Barlow, 2002) as well as in Ref. \cite {Lyons, Cowan}.
 The maximum Likelihood (ML) is used for parameter estimation. The rationale behind this method is that if the value of $l$ at its maximum for the observed data is large, the fit is good.
To estimate the relative magnitude of $l_{max-obs}$, a series of Monte Carlo data sets are  generated according to the known functional form with the fitted parameters. Then, $l_{max}$ is estimated for each of them and finally the distribution of the simulated $l_{max}$ is compared with $l_{max-obs}$ \cite{Lyons}. 
That is, the gof is estimated by

\begin{equation}
1-\alpha=1 - \int_{f(x|H_0) > f(x_{obs}|H_0)}  \! f(x|H_0) \, dx  ,    
\end{equation}

where $x$ in our case is $l_{max}$, $x_{obs}$ is the value measured in the fit to the data, and $f(x|H_0)$ is the distribution under $H_0$. 

Heinrich (2001) and Narsky (2003) show in different papers examples in which the MLV procedure fails to answer the question of how well the data are modeled by a certain density.
The strong correlation between ML estimators and MLV leads one to accept the null hypothesis, even in cases in which the null hypothesis is clearly false. In practice, MLV tests the null hypothesis distribution against another distribution from the same family, not against any other possible distribution. That is, if data are modeled as a certain Gaussian distribution, MLV gives no information about the possibility that the sample data were non-Gaussian.
In the conclusion of his work, after performing a series of tests, Heinrich states: ``My suspicion is that MLV never works (that is, it can't discover if the data does not match the form of the fitted distribution), but in any case, I would not use the method for a particular distribution without proof of validity `` \cite{Heinrich}. Narsky  and Lyons'(2008) conclusions are similar \cite {Narsky, Lyons2, Narsky2}.

\subsection{ Nearest Neighbors }
\label{sec:4}

Another popular gof procedure is to use nearest neighbors. Unlike $\chi^2$, and in general the majority of other gof techniques, methods based on Euclidean distance between nearest measured events are not greatly affected by the ``curse of dimensionality'' \cite{Breiman}. That is, they can be more easily applied to an arbitrary number of dimensions.
A first model  to test gof using nearest neighbors has been suggested by Clark and Evans in 1954 for bivariate distributions \cite{1954} and by the same authors in 1979 for any dimension \cite{1979}.
They propose a test based on the average distance between nearest neighbors in a region of space $A$.

Diggle (1979) points out that Clark and Evans' approach  ignores the inherent dependencies  among the distances, which can lead to wrongly rejecting the null hypothesis. Diggle proposes a different nearest neighbor gof model that matches the marginal distribution function of nearest neighbors with the EDF. That is, one applies a univariate gof test such as Kolmogorov-Smirnov or Cramer-von Mises using as an EDF an entire distribution of ordered distances to nearest neighbors\cite{Diggle}.

Ripley (1977) proposes to use a function $K(t)$ that gives the number of points within distance $t$ from a given point. The gof is tested by estimating the maximum difference between the measured and the expected value of $K(t)$.

Bickel and Breiman (1983) introduce a multidimensional gof test that considers the distribution of the variable

\begin{equation}
{W_i}= exp \{-ng(X_i)V(R_i)\},    
\end{equation}    

where $g(X_i)$ is the hypothesized density at the point $X_i$, $R_i$ is the distance from $X_i$ to its nearest neighbor, and $V(r)$ is the volume of a nearest neighbor sphere centered at $X_i$.   Later Schilling (1983) generalizes this statistic by adding some specific weights that allow the test to better discriminate against prespecified contiguous alternatives of $g(X_i)$ \cite {Schilling}. An approximation of both these statistics is given by Schilling for applications in which the number of dimensions is extremely high \cite {Schilling2}.  Voronoi regions can be used as well instead of spheres \cite{D0}.

Narsky (2003) suggests a nearest neighbor gof test that considers minimal and maximal cluster size. The average  distance $d_i^{(m)}$ from the center of the cluster to $m$ nearest neighbors is taken as a measure of cluster size. Ref. \cite {Narsky2} gives suggestions on optimization of parameter $m$. The probability of observing the smallest and the largest cluster is used to test gof. 

A large number of nearest neighbor related procedures have been applied to two-sample multivariate testing problems, see, for example, Ref.  \cite {TwoSample}. The rationale behind two-sample tests is similar to gof tests. The two data sets are mixed together and  tests are typically based on counting the number of nearest neighbor events that belong to the same sample. 

Gof tests based on nearest neigbors are particularly suitable for localizing maximal deviations or small clusterings in the data, i.e. unexpected peaks in the data. That is, the kind of analysis in which the Kolmogorov-Smirnov performance is weak. For a comparison between Narsky's distance to nearest neighbor and  Kolmogorov-Smirnov performance in multi-dimension, see Ref. \cite{Narsky}. The nearest neighbor approach performs better for every test, showing a remarkable versatility against different alternative hypotheses. However, Narsky points out that ``by no means it should be expected to provide the best discrimination against every alternative hypothesis''. When more is known about the alternative hypothesis, other tests are likely to achieve a better performance.

\subsection{ Neyman Smooth Tests }
\label{sec:5}

Neyman's smooth gof test (1937) is historically considered to be the first smooth test of gof (although later it was proved that Pearson's $\chi^2$ was a smooth test as well).  
Unlike other tests discussed in this work, the Neyman test specifies the alternative hypothesis. That is, they define the alternative to $f(x)$ as

\begin{equation}
g(x)= C(\theta) exp [\displaystyle\sum_{i=1}^k \theta_i h_i (x) ] f(x),    
\end{equation}    

where C is a normalization function, $\theta$ are free parameters, and $h_i(x)$ are a set of orthonormal functions. If the null hypothesis is a uniform distribution, $h_i(x)$ are Legendre polynomials. For suggestions on how to choose $h_i(x)$ when checking for different distributions, see Ref. \cite{Rayner}. 
 The Gof statistic is derived from the Rao statistic, that is

\begin{equation}
 \displaystyle\sum_{i=1}^k  [\displaystyle\sum_{i=1}^n  h_i (x) ]^2 ,    
\end{equation}    

the null hypothesis is rejected for large values of the statistic.
Crucial for the performance of  Neyman smooth tests is the optimization of $k$. For a detailed overview about the possible choices of $k$, see Ledwina (1994). In her work, she suggests to use Schwarz's Bayesian information criterion  \cite{Ledwina}. Notably,  she shows how choosing the right value for $k$ dramatically improves test power.
For an example of the power of the Neyman smooth gof test applied to multivariate physics data, see Ref. \cite{AslanComparison}. The Neyman multivariate smooth test performs similarly to the Mardia's test. No indication about the optimization of the statistic parameter is given. 

\subsection{Tree/Classifier Based Approach}
\label{sec:6}

Friedman and Rafsky (1979) propose a generalization of the Wald-Wolfowitz run test based on Minimum Spanning Tree (MST). Although the MST test is theoretically applicable to any  univariate test based on an EDF,  it  achieves the best performance when used in conjunction with the Wald-Wolfowitz test.
 The Wald-Wolfowitz  univariate two-sample test sorts all events in ascending order. Then, it labels each event depending on its original sample. Defining a $run$ as a sequence of consecutive events with the same label, the null hypothesis is rejected for small run lenghts \cite{Wald}.   
 In a multivariate MST,  a tree that connects all points without allowing closed circles is built. Events connected by a link are the closest events; that is, they are the nearest neighbors. In fact, the MST test is, from a theoretical point of view,  a nearest neighbor  method. 
When performing the Wald-Wolfowitz test, links between events from different samples are  removed. The null hypothesis is rejected for small numbers of sub-trees generated by this procedure \cite{Friedman}.  
Interestingly, Friedman and Rafsky performed a series of power studies comparing the MST Wald-Wolfowitz test with a Kolmogorov-Smirnov multivariate generalization. In one dimension, the Kolmogorov-Smirnov method is well-known to have a better performance and experimental results confirm it. However, as the  number of dimensions increases, the difference between the two test performance decreases, and for number of dimensions$>5$  Wald-Wolfowitz outperforms  Kolmogorov-Smirnov. This, once again, proves that the notion of rank is hardly extensible to high dimensions \cite{Friedman}. 

In 2003, Friedman presented a test based on any possible machine learning classifier. However, Friedman suggests choosing a decision tree as a classifier \cite{FriedmanNew}. In the case of rejection of the null hypothesis, a tree allows one to visualize the region of space in which observed data and the fit model differ. The procedure when a tree is used as a classifier is briefly decribed here, however the rationale is the same for any other machine learning classifier.

Different labels (+1/-1) are assigned to data from different samples. Then, a tree is built to separate the samples. Various optimization criteria to create a tree exist,  Gini index is discussed here. The value of the Gini index associated with the original data is used to test gof. To derive the Gini index distribution, class labels are randomly permuted among the data many times and the resulting Gini indexes are estimated as for the original samples. 
The fraction of experiments that gives a negative Gini index higher than the original one is used to compute gof.  

As Friedman himself notes in his work, a potential disadvantage of this method is that a machine learning classifier is very likely to find differences between samples, given enough number of events. That is, even  small differences between observed and expected data  could lead to reject the null hypothesis \cite{FriedmanNew}.  A useful description of a practical application of Friedman's test can be found in the documentation of the statistical package StatPatternRecognition (SPR) \cite{SPR}. Advice on the optimization of classifier parameters, specifically for the gof procedure, are given in the SPR documentation as well.

\subsection{Distance Based Approach}
\label{sec:7}

Cuadras et al. (1995,1997, 2003) introduce a multivariate Distance Based (DB) gof test based on the squared distance between two distributions. 
Given two samples of data $X$ and $Y$, they define the geometric variability of $X$ as

\begin{equation}
V_d(x)= \frac {1} {2}  \int_{S^2} d^2 (x,x') f(x) f(x') \lambda(dx) \lambda(dx')  ,    
\end{equation}    

where $d(x,x')$ is a  distance function on the space $S$ and $\lambda$ a specific measure chosen according to the characteristics of  both vectors.  $V_d(X)$,  a version of Rao's quadratic entropy, is a generalization of the variance of $X$ with respect to dissimilarity $d$.

The squared distance between samples $X$ and $Y$ is defined as

\begin{equation}
\Delta^2(\Pi_1, \Pi_2)=  \int_{S^2} d^2 (x,y) f(x) g(y) \lambda(dx) \lambda(dy) - V_d(X) -V_d(Y)  ,    
\end{equation}    

where $\Pi_1$ and $\Pi_2$ are the two populations represented as samples $X$, $Y$ with density functions $f(x)$ and $g(y)$. $\Delta^2$ is a Jensen difference  that can be interpreted as a distance between two mean vectors \cite{Rao,Cuadras2}.

The sampling version of  DB test is then:

\begin{equation}
\hat{V}_d(X)= \frac{1}{2n_1^2} \displaystyle\sum_{i=1}^{n_1}  \displaystyle\sum_{j=1}^{n_1} d^2(x_i,x_j)  ,    
\end{equation}    

and

\begin{equation}
\hat{\Delta}^2(\Pi_1, \Pi_2)= \frac{1}{n_1n_2} \displaystyle\sum_{i=1}^{n_1}  \displaystyle\sum_{j=1}^{n_2} d^2(x_i,y_i)-\hat{V}_d(X)  -\hat{V}_d(y)   .    
\end{equation}    

Obviously, small values of the squared distance indicate strong similarity between the two samples. In fact, DB tests the null hypothesis $\Delta^2=0$.  
In Ref. \cite{Cuadras2}, several applications of the  DB test are presented, highlighting its versatility.

Independently from Cuadras et al., Zech and Aslan (2003-2005) suggest a multivariate test that can be viewed as a generalization of the DB one. They propose the Energy test whose two-sample statistic is

 \begin{equation}
\Phi_{n_1n_2}= \frac{1}{n_1^2} \displaystyle\sum_{i<j}^{n_1} R(|x_i-x_j|) +\frac{1}{n_2^2} \displaystyle\sum_{i<j}^{n_2} R(|y_i-y_j|) + \frac{1}{n_1n_2} \displaystyle\sum_{i=1}^{n_1} \displaystyle\sum_{j=1}^{n_2} R(|x_i-y_j|) ,
\end{equation}    

where $R(r)$ is a continuous, monotonic function of the Euclidean distance between the vectors. They propose different functions for $R(r)$. The best function depends on the characteristics of the data. However, they find that $R=- ln (r+\epsilon)$ is generally the best choice, offering a good rejection power against many different alternatives. Here, $\epsilon$ is a cut-off introduced to improve the power of the test by avoiding problems caused by extremely small values of the distance. For a detailed description of possible distance functions as well as for a mathematical definition of the cut-off, see Ref. \cite{Aslan} .
Thus, Zach and Aslan suggest to compare two samples by using a logarithmic measure of point to point dissimilarity.

Remarkably, Zach and Aslan conducted extensive studies about  power of their test in different configurations using physics data. In one dimension, the energy test is competitive with the most powerful univariate tests. In multidimensions, the energy test performs significantly better than any other test it has been compared with: Mardia's test, Neyman smooth test,  a nearest neighbors test and Friedman-Rafsky test. Specifically, the best relative performance is achieved in the highest dimension (D=4).

Lyons (2008), after reviewing the weaknesses of EDF and $\chi^2$ tests in multidimension, seems to consider DB/Energy tests as the most reliable choice in high dimensions.

\subsection{Summary}
\label{sec:8}

Univariate gof methods have been extensively studied and analyzed. In high dimensions, a  significant number of gof tests for unbinned data have been proposed. However, an accurate study of their power, in different configurations, has never been performed.
Ideally, one would look for a test that is uniformly most powerful. In practice, such a test does not exist.The test power depends on the specific problem. Therefore, the choice of the test should depend on the analysis characteristics. 
Nearest neighbor tests seem a good choice when the alternative hypothesis is composite. Energy/DB tests have been shown to achieve good results in high dimensions. However, there is no ultimate ``best'' gof test. It would be very useful to check different gof tests on several different alternatives in order to gain an understanding of the circumstances in which a specific gof test works well and those when it does not.

\section*{Acknowledgement}
Thanks to Bob Cousins for insightful suggestions, engaging in fruitful discussions, and reviewing earlier drafts. Thanks to Julian Bunn for reviewing the final draft of this work.

\end{document}